\documentclass[10pt ]{article}

\usepackage{amsmath,amsthm,amsfonts,amssymb}

\chardef\bslash= `\\

\theoremstyle{definition}
\newtheorem{defn}{Definition}[section]

\theoremstyle{remark}
\newtheorem{rem}{\bf Remark}[section]

\theoremstyle{definition}
\newtheorem{com}{Comment}[section]

\newcommand{\eval}[2][\right]{\relax
  \ifx#1\right\relax \left.\fi#2#1\rvert}

\begin{document}

\noindent
{\bf Alexander I. Nesterov, Lev V. Sabinin}
\vspace*{0.05truein}

\noindent
{\Large \bf \em  Non-associative geometry and discrete
structure of spacetime }
\vspace*{0.035truein}

\vspace*{0.21truein}

\noindent
{\bf Abstract:} A new mathematical theory, {\em non-associative geometry},
providing a unified algebraic description of continuous and discrete
spacetime, is introduced.

~~


\section{Introduction}

The recent development of geometry has shown the importance
of non-associative algebraic structures such as quasigroups, loops and
odules. For instance, it is possible to say that the non associativity is
the algebraic equivalent of the differential geometric concept of
curvature. The corresponding construction may be described as follows. In
a neighborhood of an arbitrary point on a manifold with an affine
connection one can introduce the geodesic local loop which is uniquely
defined by means of the parallel translation of geodesics along geodesics
\cite{K,S1,S2}. The family of local loops constructed in this way uniquely
defines a space with affine connection, but not every family of geodesic
loops on a manifold defines an affine connection. It is necessary to add
some algebraic identities connecting loops in different points. Later, the
additional algebraic structures (so called {\em geoodular structures}) were
introduced and the equivalence of the categories of geoodular structures
and of affine connections was shown by Sabinin \cite{S3,S4}. The main
algebraic structures arising in this approach are related to non
associative algebra and theory of quasigroups and loops.

In our paper the new mathematical theory, {\em non-associative geometry},
which may help us to understand the discrete structure of spacetime, is
introduced. The key point is to give an algebraic (non-local) description
of manifold, which may be used in continuous and discrete cases.
Non-associative geometry provides the unified algebraic description of
continuous and discrete spacetime and admits all basic attributes of
spacetime geometry including generalized Einstein's equations.

\section{What is the non-associative geometry?}

Here we survey, in brief, algebraic foundations of {\em non-associative
geometry} due to L.V. Sabinin (see on the matter \cite{S5,S6,S7,S8} ).

\defn Let $\langle Q,{\boldsymbol\cdot}\rangle $ be a groupoid with a
binary operation $(a,b) \mapsto a \boldsymbol\cdot b$ and $Q$ be a smooth
manifold. Then $\langle Q,{\boldsymbol\cdot}\rangle $ is called a
{\it quasigroup} if the  equations $a{\boldsymbol\cdot}
x=b,~y{\boldsymbol\cdot} a=b$ have a unique solutions:  $x=a\backslash b$,
$y=b/a$. A {\it loop} is a quasigroup with a two-sided identity,
$a{\boldsymbol\cdot} e= e{\boldsymbol\cdot} a=a, \forall a \in Q$. A loop
$\langle Q,{\boldsymbol\cdot},e \rangle$ with a smooth functions
$\phi(a,b):=a{\boldsymbol\cdot} b$ is called a {\it smooth loop}.

Let $\langle Q,{\boldsymbol\cdot},e\rangle$ be a smooth local loop with a neutral element $e$.
We define
\begin{equation}
L_a b=R_b a=a{\boldsymbol\cdot} b,\quad
l_{(a,b)}=L^{-1}_{a{\boldsymbol\cdot} b}\circ L_a\circ L_b,
\label{Ll}
\end{equation}
where $L_a$ is a {\it left translation}, $R_b$ is a {\it right
translation}, $l_{(a,b)}$ is an {\it associator}.

\defn{Let $\langle{ M},\cdot,e \rangle$ be a partial {\it
groupoid} with a binary operation $(x,y)\mapsto x\cdot y$ and a
neutral element $e ,\; x\cdot e  =e \cdot
x =x$; $   M$ be a smooth manifold (at least $C^1$-smooth) and
the operation of multiplication (at least $C^1$-smooth) be defined
in some neighborhood $U_e $,} then  $\langle {
M},\cdot,e \rangle$ is called a {\it partial loop on
$   M$.}

\rem{The operation of multiplication is locally left and right
invertible. This means that  if $x\cdot y = L_x y= R_y x$ then there exist
$L^{-1}_x$ and $R^{-1}_x$ in some neighborhood of the neutral element
$e $:
\[
L_a(L^{-1}_a x)=x, \quad R_a(R^{-1}_a x)=x.
\]
}
{The vector fields $A_j$ defined on $U_e $ by
\begin{equation}
A_j(x)=\bigl((L_x)_{*,e }\bigr)^i_j\frac{\partial}{\partial x^i} =
L^i_j(x)\frac{\partial}{\partial x^i}
\end{equation}
are called the {\it left basic fundamental fields}. Similarly, the
{\it right basic fundamental fields} $B_j$ are defined by
\begin{equation}
B_j(x)=\bigl((R_x)_{*,e }\bigr)^i_j\frac{\partial}{\partial
x^i} = R^i_j(x)\frac{\partial}{\partial x^i}.
\end{equation}
}
 The solution of the equation
\begin{equation}
\frac{df^i(t)}{dt} =L^i_j(f(t))X^j, \quad f(0) = e,
\end{equation}
is of the form $f(t)={\rm Exp} tX$ defining
 the {\it exponential map}
\[
{\rm Exp}: X\in T_e  ({   M})\longrightarrow
{\rm Exp}X \in{   M}.
\]
The unary operation
\begin{equation}
tx={\rm Exp}(t {\rm Exp}^{-1}x),
\end{equation}
based on the exponential map, is called the {\it left canonical
unary operation} for $\langle {M},\cdot,e \rangle$.
A smooth loop $\langle {M},\cdot,e \rangle$
equipped with its canonical left unary operations is called the
{\it left canonical preodule} $\langle {
M},\cdot,(t)_{t\in\mathbb  R}, e \rangle$. If one more operation is
introduced,
\begin{equation} x+y ={\rm Exp}({\rm Exp}^{-1}x + {\rm Exp}^{-1}y),
\end{equation}
then we obtain the {\it canonical left prediodule} of a loop,
$\langle{   M},\cdot,+,(t)_{t\in\mathbb  R}, e \rangle $.
 {A canonical left
preodule (prediodule) is called the {\it left odule (diodule)} if the {\it
monoassociativity} property
\begin{equation}
tx\cdot ux = (t+u)x
\end{equation}
is satisfied.} In the smooth case, for an odule, the left and the right
canonical operations as well as the exponential maps coincide.

\defn{Let $   M$} be a smooth manifold and
\[
L:  (x,y,z)\in {   M} \mapsto L(x,y,z)\in {   M}
\]
a smooth partial ternary operation, such that $x_{\dot a}y =L(x,a,z)$
defines  in  some neighbourhood of the point $a$   the loop with the
neutral $a$, then the pair $\langle {   M}, L\rangle$  is called a {\it
loopuscular structure (manifold)}.

{A smooth manifold  $   M$ with a smooth partial ternary operation
$L$ and smooth binary operations $\omega_t : (a,b)\in {   M}\times{
M} \mapsto \omega_t(a,b)=t_a b\in {   M}, \;(t\in \mathbb  R)$, such that
$x_{\dot a}y =L(x,a,y)$ and $t_a z = \omega_t(a,z)$ determine in some
neighborhood of an arbitrary point $a$ the odule with the neutral element
$a$, is called a {\it left odular structure (manifold)} $\langle {   M},
L, (\omega_t)_{t\in {\mathbb  R}}\rangle$. Let $\langle {   M}, L,
(\omega_t)_{t\in {\mathbb  R}}\rangle$ and $\langle {   M}, N,
(\omega_t)_{t\in {\mathbb  R}}\rangle$ be odular structures, then $\langle
{   M},L, N, (\omega_t)_{t\in {\mathbb  R}}\rangle$ } is called a
{\it diodular structure (manifold). If $x_{\stackrel{+}a}y =
N (x,a,y)$} and $t_a x =\omega_t(a,x)$ define a vector space, then
such a diodular structure is called a {\it linear diodular structure}.

A diodular structure is said to be  {\it geodiodular} if
\begin{align}
&\text{(the first geoodular identity)}\qquad
L_{u_ax}^{t_ax}\circ L_{t_ax}^{a}=L_{u_ax}^{a}\quad (L_x^ay=L(x,a,y)),\\
&\text{(the second geoodular identity)} \quad
L_{x}^{a}\circ  {t_a}=t_x \circ  L_{x}^{a},\\
&\text{(the third geoodular identity)} \quad \; \;
L_{x}^{a}N(y,a,z)= N(L_{x}^{a}y, {x},
L_{x}^{a}z)
\end{align}
are true.

\defn {Let $   M$ be a $C^k$-smooth $(k\geq 3)$
affinely connected manifold and the following operations are given on
$ M$ (locally):
\begin{align}
& L^a_x y= {x\; {}_{\stackrel{\cdot}{a}}\; y}
={\rm Exp}_x \tau^a_x {\rm Exp}^{-1}_a y, \\
& \omega_t(a,z) = t_a z= {\rm Exp}_a t {\rm Exp}^{-1}_a z, \\
&N (x,a,y) = {x\; {}_{\stackrel{+}{a^{}}}\; y}
={\rm Exp}_a ({\rm Exp}^{-1}_a x + {\rm Exp}^{-1}_a y),
\end{align}
${\rm Exp}_x$ being the exponential map at the point $x$ and $\tau^a_x$
the parallel translation along the (unique) geodesic going from $a$ to $x$.
The above construction equips $M$ with the linear geodiodular structure
which is called a {\em natural linear geodiodular structure of an affinely
connected manifold} $({M},\nabla)$.}

\rem {Any $C^k$-smooth $(k\geq 3)$  affinely connected manifold can be
considered as a geodiodular structure.}

\defn {Let $\langle {   M}, L\rangle$ be a
loopuscular structure of a smooth manifold $   M$. Then the formula
\begin{gather*}
\nabla_{X_a}Y=\left\{\frac{d}{dt}
\left([(L^a_{g(t)})_{*,a}]^{-1}Y_{g(t)}\right)\right\}_{t=0},\\
g(0)=a, \quad \dot g(0)=X_a,
\end{gather*}
$Y$ being a vector field in the neighborhood of a point $a$, defines the
{\em tangent affine connection}.}

In coordinates the components of this affine connection are
\[
\Gamma^i_{jk}(a) = - \left[\frac{\partial^2 (x{}_{\dot a}{}y)^i}{\partial
x^j \partial y^k}\right]_{x=y=a}.
\]
The equivalence of the categories of geoodular (geodiodular) structures and
of affine connections has been shown in \cite{S3,S4}.

\defn{Let $\langle M,L\rangle$ be a loopuscular structure, then
\begin{equation}
h^a_{(b,c)} = (L^a_c)^{-1}\circ L^b_c \circ L^a_b
\label{hol}
\end{equation}
is called the {\em elementary holonomy}.}

\com{The elementary holonomy is, in fact, the parallel translation along
a geodesic triangle path. Consequently, it is some integral curvature.
Indeed, in the smooth case, differentiating $(h^a_{(x,y)})^i$ by
$x^j,\, y^k$ at $a\in M$ we get the curvature tensor at $a\in M$  precisely
up to numerical factor,
\begin{equation}
R^i{}_{jkl}(a)= 2 \left[\frac{\partial^3 (h^a_{(x,y)} z)^i}
{\partial x^l\partial y^k\partial z^j}\right]_{x=y=z=a}
\label{hl}
\end{equation}
}

\com{For a diodular structure one can consider an elementary holonomy
$h_{(a, b)}= h_{(a, b)}^e$ together with the diodule
$\langle{   M},\; {}_{\stackrel{\cdot}{e}}\;,\;
{}_{\stackrel{+}{e^{}}}\;,(t_e)_{t\in\mathbb  R}, e \rangle$, so called
{\it holonomial diodule}, and restore this diodular structure in a unique
way:
\begin{align}
& L (x, a, y)= L_x^e h_{(a, x)}(L_a^e)^{-1}y,
\\
&N (x, a, y)= L_a^e ( (L_a^e)^{-1}x\; {}_{\stackrel{+}{e^{}}}\;(L_a^e)^{-1}y),
\\
& \omega_t(a,y) = t_a y=L_a^e t_e (L_a^e)^{-1}y.
\end{align}
In this case the  holonomial identities
\begin{align}
& h_{(a, b)}t_e x=t_eh_{(a, b)}x,\quad
h_{(a, b)}(x\; {}_{\stackrel{+}{e^{}}}\;y)=
h_{(a, b)}x\; {}_{\stackrel{+}{e^{}}}\;
h_{(a, b)}y
\quad\text{(linearity),}
\label{Hol1}\\
& h_{(a, a\boldsymbol\cdot u_eb)}tb=l_{(a, u_eb)}tb\quad\text{(joint identity)},
\\
&h_{(c\boldsymbol\cdot t_ea, c\boldsymbol\cdot u_ea)}
h_{(c, c\boldsymbol\cdot t_ea)}x=h_{(c, c\boldsymbol\cdot u_ea)}x
\quad\text{($h$-identity),}
\\
&h_{(e, q)}x=x \quad \text{($e$-identity)}
\label{Hol2}
\end{align}
are true.}

\com{Using the definition of  elementary holonomy (\ref{hol}) we may easily verify that elementary holonomy satisfies the odular Bianchi identities:
\begin{equation}
h^a_{(z,x)}\circ
h^a_{(y,z)}\circ h^a_{(x,y)} = (L^a_x)^{-1}\circ h^x_{(y,z)} \circ L^a_x
\label{B1} .
\end{equation}
These identities can be considered as non-local form of the usual Bianchi
identities and in this sense they are equivalent. Perhaps,
this is the first non-local algebraic expression of the Bianchi
identities. In the linear approximation (\ref{B1}) generates  the
Bianchi identities in conventional  form. This may be easily seen
in  the normal coordinates related to the point $a$.}

The non-associative geometry is based on the constructions
described above. In the table below we compare the  basic concepts of the
classical differential geometry and of the non-associative geometry.

\begin{center}
{\bf Differential Geometry vs Non-associative
Geometry}
\end{center}
{
\begin{center}
\begin{tabular}{|l|l|}\hline \hline
Differential Geometry & Non-associative Geometry \\ \hline \hline

Tangent space $T_a(M)$ & Osculating space
$\langle M, \underset{a}{+}, \,a\,, (t_a)_{t\in \mathbb R}\rangle$ \\ \hline

Tangent bundle structure   & Osculating structure
$\langle M, N, (\omega_t)_{t\in \mathbb R}\rangle$\\ \hline

Cotangent space & Co-osculating space \\ \hline

Parallel displacement &
Left translations $L^a_x y$  \\ \hline

Curvature $R(X,Y)Z$& Elementary holonomy $h^a_{(x,y)}z$\\  \hline

Bianchi identities& Odular Bianchi identities\\ \hline

\end{tabular}
\end{center} }

\subsection{Example: the non-asociative geometry of two-dimensional sphere
$S^2_{R}$}

The well known two-sphere $S^2_{R}$ of radius  $R$ admits a natural loop structure
\cite{N0,N1,N2} which may be described as follows. Let ${\mathbb C}$ be a
complex plane and $\zeta,\eta\in{\mathbb C}$. The non-associative
multiplication $\star$ is defined by
\begin{equation}
\zeta\star\eta =L_\zeta \eta=\frac{\zeta+\eta}{1-
{\overline\zeta\eta}/{R^2}}, \quad \zeta,\eta\in{\mathbb C} \label{Lp}
\end{equation}
where bar denotes complex conjugation and the neutral element $e$ coincides
with the origin of system of coordinates. This loop is isomorphic to the
local two-parametric loop associated with two-sphere $S^2_{R}$.  The
isomorphism between points of the sphere and points of the complex plane
${\mathbb C}$ is established by the stereographic projection from the south
pole of the unit sphere, $\zeta = R\tan(\theta/2)e^{\rm i\varphi}$.

\rem {The entire sphere may be covered by two local (partial) loops, one of
them with the neutral element at the north pole (see the above) and another
with the neutral element at the south pole.}

The associator is found to be
\begin{equation}
l_{(\zeta,\eta)}\xi=\frac{1-\zeta\overline\eta/R^2}
{1-\eta\overline\zeta/R^2 }\xi.
\end{equation}
The sphere is a symmetric space and its elementary holonomy is determined
by the associator: $h_{(\zeta,\eta)} =l_{(\zeta,L^{-1}_\zeta\eta)}$
\cite{S5,S8}.  The computation gives
\begin{equation}
h_{(\zeta,\eta)}\xi=\frac{1+ \overline\zeta \eta /R^2}
{1+\zeta\overline\eta/R^2}\xi.
\label{hol1}
\end{equation}

The left invariant diodular metric on $S^2_{R}$ is given by
\begin{equation}
g^0(L^{-1}_\zeta\xi,L^{-1}_\zeta\eta) =  g^{\zeta}(\xi,\eta),
\label{dmetr}
\end{equation}
where $g^0(\zeta,\eta)$ is the diodular metric tensor at the neutral
element induced by the natural metric on the tangent space at the
neutral element (north pole of $S^2_{R}$), and $g^{\zeta}(\xi,\eta)$ is the
diodular metric at the point $\zeta$. Actually (\ref{dmetr}) is an
algebraic analogue of compatibility of the connection with the metric
structure of $S^2_{R}$. We define the left invariant diodular metric on
two-sphere as follows:
\begin{equation}
g^{\zeta}(\xi,\eta) = 2\left(\frac{(\xi-\zeta)(\bar\eta - \bar\zeta)}
{(1 + \bar\zeta\xi/R^2)(1+ \zeta\bar\eta/R^2)}
+ \frac{(\bar\xi - \bar\zeta)(\eta - \zeta)}
{(1 + \zeta\bar\xi/R^2)(1+ \bar\zeta\eta/R^2)}\right) .
\nonumber
\end{equation}
In particular,
\begin{equation}
g^{\zeta}(\xi,\xi) = \frac{4|\xi-\zeta|^2}{|1 +
\bar\zeta\xi/R^2|^2 } .
\label{dm}
\end{equation}
Let $\xi = \zeta + d \zeta $, then (\ref{dm}) leads to
\[
g(d\zeta,d\zeta) = \frac{4d\zeta d\bar\zeta}{(1 + |\zeta|^2/ R^2)^2},
\]
the well known expression for the element of length of $S^2_{R}$.

{\com Note that the same result may be obtained in another way.
Let us introduce the basis of left fundamental vectors and the  dual basis
of one-forms:
\begin{eqnarray}
\Gamma_1 = (1 + |\zeta|^2)\partial_\zeta,
\quad \Gamma_2 = (1 + |\zeta|^2)\partial_{\bar\zeta} \\ \theta^1 =
\frac{d\zeta}{1 + |\zeta|^2}, \quad \theta^2 = \frac{d\bar\zeta}{1 +
|\zeta|^2}.
\end{eqnarray}
Then the metric based on the left fundamental basis forms is given by
\begin{equation}
ds^2 = 4 \theta^1 \theta^2 =
\frac{4d\zeta d\bar\zeta} {(1 + |\zeta|^2/R^2)^2}.  \label{int}
\end{equation}
Computation of the curvature tensor gives
\begin{equation}
R^\zeta{}_{\zeta\zeta \bar\zeta} = \frac{2}{R^2 \left(1 + {|\zeta|^2}/{
R^2}\right)^2},
\label{curv}
\end{equation}
and using (\ref{hol1}) we find
\[
 R^\zeta{}_{\zeta\bar\zeta \zeta}(0)
=  2\left[\frac{\partial^3 (h_{(\zeta,\eta)}\xi)}
{\partial\zeta\partial\bar\eta \partial \xi}\right]_{\zeta=\eta=\xi=0 } ,
\]
which is consistent with (\ref{hl}).
}

\subsubsection{The non-associative discrete geometry of $S^2_{R}$ }

We start with the natural geodesic triangulation of the sphere which is
specified as follows. The simplex, triangulating $S^2_{R},$ is a geodesic
triangle.  The geodesic lattice will be assumed to consist of central
vertex at the north pole and of the geodesic triangles attached to this
vertex. To each surface vertex $ {\bf p}=(j,k) $ we assign the polar
coordinates $(\theta_j,\varphi_k)$, assuming $\theta_j= \pi j/n ,
\;\varphi_k = 2\pi k/n \;(j,k=0,1,2,\dots n-1)$.  With such a choice we
have the triangulation defined by $n^2$ points allocated on the suraface of
the sphere:
\begin{equation}
\zeta_{\bf p}=R \tan\biggl(\frac{\pi j}{2n}\biggr) e^{\frac{2\pi i k}{n}}.
\label{Tr}
\end{equation}

The non-associative operation (\ref{Lp}) now takes the form
\begin{equation}
\zeta_{\bf pq} =\frac{\zeta_{\bf p}+\zeta_{\bf q}}{1-\overline\zeta_{\bf
p}\zeta_{\bf q}/R^2}.
\label{DLp}
\end{equation}
Writing $\zeta_{{\bf p}{\bf q}}$ as
\begin{equation}
\zeta_{{\bf p}{\bf q}}= R \tan\biggl(\frac{\theta_{{\bf p}{\bf
q}}}{2}\biggr) e^{i\varphi_{{\bf p}{\bf q}}}, \label{DLp2}
\end{equation}
one obtains from (\ref{DLp})   the following formulae which define the left
translations in ``spherical coordinates'':
\begin{align}
&\theta_{{\bf p}{\bf q}} =2
\tan^{-1}\biggl(\frac{1}{R}\biggl|\frac{\zeta_{\bf p} + \zeta_{\bf q}}
{1 - \bar\zeta_{\bf p}\zeta_{\bf q}/R^2}\biggr|\biggr) \\
&\varphi_{{\bf p}{\bf q}} = \arg(\zeta_{\bf p} + \zeta_{\bf q})-\frac{i}{2}\ln
l{(\zeta_{\bf p},\zeta_{\bf q})}, \\
&\zeta_{{\bf p}}= R \tan\biggl(\frac{\pi { j}}{2n}\biggr)
e^{\frac{2\pi i {k}}{n}}, \qquad \zeta_{{\bf q}}= R
\tan\biggl(\frac{\pi {l}}{2n}\biggr) e^{\frac{2\pi i {m}}{n}},
\end{align}
where
\[
l{(\zeta_{\bf p},\zeta_{\bf q})} = \frac{1-\zeta_{\bf
p}\overline\zeta_{\bf q}/R^2} {1-\overline\zeta_{\bf p}\zeta_{\bf q}/R^2} .
\]
For the duiodular metric we have
\begin{equation}
g^{\zeta_{\bf p}}(\zeta_{\bf q},\zeta_{\bf q}) =
\frac{4|\zeta_{\bf p}-\zeta_{\bf q}|^2}
{|1 + \overline\zeta_{\bf p}\zeta_{\bf q}/R^2|^2 },
\label{dm2}
\end{equation}
and the elementary holonomy is
\begin{equation}
h_{(\zeta_{\bf p},\zeta_{\bf q})}\zeta_{\bf m}
=\frac{1+ \overline\zeta_{\bf p}\zeta_{\bf q} /R^2}
{1+ \zeta_{\bf p}\overline\zeta_{\bf q} /R^2} \zeta_{\bf m}.
\label{hol2}
\end{equation}
\com{To certain extent the information concerning the geometry of the
sphere is hidden in the structure of the finite loop. The spherical
symmetry is determined by the relation between the associator and
elementary holonomy, $h_{(\zeta_{\bf p},\zeta_{\bf q})} =l_{(\zeta_{\bf
p},L^{-1}_{\zeta_{\bf p}} \zeta_{\bf q})}$. The smooth sphere could be
regarded as the result of ``limit process'' of triangulating, while $n\longrightarrow
\infty$. Indeed, with the growing of $n$, the number of the points
increases and the triangulations become more fine.
}

In order to obtain the correct `continuous' limit let us consider ${\bf q = p
+ \mbox{\boldmath$\delta$}}, |\mbox{\boldmath$\delta$} | \ll n$. Let
$\mbox{\boldmath$\delta$}= (l,m)$, then
\[
\zeta_{\bf q} = \zeta_{\bf p} +
R\left(\frac{\zeta_{\bf p}}{ \overline\zeta_{\bf p}}\right)^{\frac{1}{2}}
\left(\Bigl(1 + \frac{|\zeta_{\bf p} |^2}{R^2}\Bigr) \frac{\pi l}{2n}
+ i\frac{|\zeta_{\bf p}|}{R} \frac{2\pi m}{n}\right)  +
O \left(\biggl(\frac{|\mbox{\boldmath$\delta$}|}{n}\biggr)^2 \right),
\]
and the diodular metric takes the form
\begin{equation}
g^{\zeta_{\bf p}}(\zeta_{\bf q},\zeta_{\bf q}) =
R^2\left(\Bigl(\frac{\pi l}{n}\Bigr)^2 +
\frac{4|\zeta_{\bf p}|^2}{(1 + |\zeta_{\bf p}|^2/R^2)^2 }
\Bigl(\frac{2\pi m}{n}\Bigr)^2 \right) +
O \left(\biggl(\frac{|\mbox{\boldmath$\delta$}|}{n}\biggr)^2 \right) .
\label{dm3}
\end{equation}
Simplifying (\ref{dm3}), we obtain
\begin{equation}
g^{\zeta_{\bf p}}(\zeta_{\bf q},\zeta_{\bf q}) =
R^2\left( (\Delta \theta_{\bf q})^2 +  \sin^2\theta_{\bf p} (\Delta
\varphi_{\bf q})^2 \right)+
O \left(\biggl(\frac{|\mbox{\boldmath$\delta$}|}{n}\biggr)^2 \right),
\label{dm4}
\end{equation}
where $\Delta \theta_{\bf q} = \pi l/n$ and
$\Delta \varphi_{\bf q} = 2\pi m/n$. The
`differential geometry' appears as the result of  limit process while $n \longrightarrow
\infty$:
\begin{eqnarray}
&&\Delta \theta_{\bf q}  \longrightarrow
d\theta, \quad \Delta \varphi_{\bf q}  \longrightarrow d\varphi, \nonumber
\\ &&g^{\zeta_{\bf p}}(\zeta_{\bf q},\zeta_{\bf q}) \longrightarrow ds^2  =
R^2\left( (d \theta)^2 +  \sin^2\theta (d \varphi)^2 \right) .
\end{eqnarray}
The similar consideration of the elementary holonomy gives
\begin{equation}
h_{(\zeta_{\bf p},\zeta_{\bf q})}\zeta_{\bf m}
=\zeta_{\bf m}\Biggl(1 +i\frac{\Delta(\zeta_{\bf p},\zeta_{\bf q})}{ R^2}
+ O \left(\biggl(\frac{|\mbox{\boldmath$\delta$}|}{n}\biggr)^2
\right)\Biggr),
\label{hol3}
\end{equation}
where
\[
\Delta(\zeta_{\bf p},\zeta_{\bf q})= \frac{2|\zeta_{\bf p}|^2
\Delta\varphi_{\bf q}}{1 + |\zeta_{\bf p}|^2/R^2}
\]
is the area of the geodesic triangle $(e,\bf p, q)$. Approaching the limit, while
$n\longrightarrow \infty$, one restores the conventional  scalar
curvature  $1/R^2$.

\section{The algebraic generalization of vacuum Einstein's equations}

The Einstein's equations in the vacuum
\begin{equation}
G_{\mu\nu}:= R_{\mu\nu} - \frac{1}{2} g_{\mu\nu}R=0
\label{R1}
\end{equation}
can be rewritten in the tetrad basis as follows \cite{EGH}:
\label{en1}
\begin{equation}
\ast R_{abcd}\theta^b\wedge\theta^c\wedge\theta^d =0,
\label{E2}
\end{equation}
where
\[
\ast R_{abcd} = \frac{1}{2}\epsilon_{abmn}R^{mn}{}_{cd}
\]
is the dual to the Riemann tensor (latin indices $a,b,c,d$ are used
for the tetrad basis and are running over 0,1,2,3). Equations
(\ref{R1}) (or (\ref{E2})) mean:
\begin{align}
& \{00\} \; \Longrightarrow\quad R^{1}{}_{212} + R^{2}{}_{323} +R^{3}{}_{131} =0,
\nonumber \\
& \{11\} \; \Longrightarrow\quad R^{0}{}_{202} + R^{2}{}_{323}
+R^{3}{}_{030} =0,   \nonumber \\
& \{22\} \; \Longrightarrow\quad R^{0}{}_{101} + R^{1}{}_{313}
+R^{3}{}_{030} =0,   \nonumber \\
& \{33\} \; \Longrightarrow\quad R^{0}{}_{101} + R^{1}{}_{212}
+R^{2}{}_{020} =0,   \nonumber \\
& \{01\} \; \Longrightarrow\quad R^{0}{}_{221} + R^{0}{}_{331}= 0,
\nonumber \\
& \{02\} \; \Longrightarrow\quad R^{0}{}_{112} + R^{0}{}_{332}= 0,
\nonumber \\
& \{03\} \; \Longrightarrow\quad R^{0}{}_{113} + R^{0}{}_{223}= 0,
\nonumber \\
& \{12\} \; \Longrightarrow\quad R^{1}{}_{002} + R^{1}{}_{332}= 0,
\nonumber \\
& \{13\} \; \Longrightarrow\quad R^{1}{}_{003} + R^{1}{}_{223}= 0,
\nonumber \\
& \{23\} \; \Longrightarrow\quad R^{2}{}_{003} + R^{2}{}_{113}= 0,
\nonumber
\end{align}
and  may be written in the following form:
\begin{equation}
\ast R(X,Y)Z  + \ast R(Y,Z)X + \ast R(Z,X)Y =0,
\label{A1}
\end{equation}
where $X,Y,Z \in T(M)$.

Let us consider the following algebgraic equation:
\begin{align}
\ast h^e_{(x,y)}z\; {}_{\stackrel{\displaystyle +}{e}}\; \ast h^e_{(y,z)}x
\; {}_{\stackrel{\displaystyle +}{e}}\; \ast h^e_{(z,x)}y =e\quad
(\forall e,x,y,z),
\label{A2}
\end{align}
where
$$
\bigl(\ast h^e_{(x,y)}\bigr)^a_b =\frac{1}{2}g^{ac}{\epsilon_{cbmn}}g^{nl}
\bigl(h^e_{(x,y)}\bigr)_l^{m}
$$
is the dual elementary holonomy. Employing the normal coordinates with the
origin at the point $e$ and the relation $(\ref{hl})$ between the
curvature and elementary holonomy, we find that in the first approximation
$(\ref{A2})$ restores the Einstein equations $(\ref{A1})$.

We propose the algebraic system (\ref{A2}) together with
(\ref{Hol1})--(\ref{Hol2}) as a non-local generalization of the vacuum
Einstein's equations (diodular Einstein's equations). They should be
considered as the equations for constructing of the diodule at the point
$e$ (which uniquely defines the corresponding diodular space).

\com{The relation between odular Einstein's equations considered in the
neighbourhoods of the points $e$ and $a$ is established by means of
the odular Bianchi identities (\ref{B1}):
\begin{equation}
\ast h^a_{(x,y)}= \ast\bigl( L^e_a \circ h^e_{(y,a)}\circ h^e_{(x,y)}\circ
h^e_{(a,x)}\circ (L^e_a)^{-1}\bigr).
\label{B2}
\end{equation} }

In the metric gravitational theory the odular Einstein's equations
should be considered together with
\begin{equation}
g^e(x,y)=g^a(L_a^e x,L_a^ey)
\end{equation}
which relates conecction and metrics (Here $g^a(p,q)$ is a metric tensor at
$a\in M$).

In the normal coordinates with the  origin at the point $e$ we get the
complete system of the diodular Einstein's equations in the form:
\begin{align}
&\bigl(\ast h^e_{(x,y)}\bigr)^a_b z^b  +
\bigl (\ast h^e_{(y,z)}\bigr )^a_b x^b  +
\bigl (\ast h^e_{(z,x)}\bigr )^a_b y^b  = 0
\label{A3} ,\\
&\bigl (h^e_{(x_b,x_c)}\bigr )^d_f = \bigl ((L^e_{x_c})^{-1}\bigr )^d_g
\bigl (L^{x_b}_{x_c}\bigr )^g_h \bigl ( L^e_{x_b} \bigr )^h_f.
\label{A4}
\end{align}

\section{Concluding remarks}

In our paper we proposed some new {\em non-associative} approach to the
classical and discrete structure of manifolds which gives the unified
description of continuous and discrete spacetime. This means that at the
Planck scales the standard concept of spacetime might be replaced  by the
diodular discrete structure which at large spacetime scales `looks like' a
differentiable manifold.

{\bf Alexander I. Nesterov:} {Departamento de F\'\i sica, C.U.C.E.I.,
Universidad de Guadalajara, Blvd. M. Garc\'\i a Barrag\'an y Calz. Ol\'\i
mpica, Guadalajara, Jalisco, C.P. 44460, M\'exico  and Institute of Physics,
Siberian Branch Russian Academy of Sciences, Akademgorodok, 660036,
Krasnoyarsk, Russia. {\em E-mail}: nesterov@udgserv.cencar.udg.mx}
~~~~~~~\\

{\bf Lev V.  Sabinin:} {Departmento de Matem\'aticas, Universidad de
Quintana Roo, Blvd. Bahia S/N, C.P. 77019, Chetumal, Quintana Roo,
M\'exico and Russian Friendship University, Moscow, Russia.
{\em E-mail}:  lsabinin@balam.cuc.uqroo.mx }

\end{document}